\documentclass[aps,prd,onecolumn,nofootinbib,superscriptaddress]{revtex4}
\pdfoutput=1
\usepackage{graphicx}
\usepackage{amsmath}
\usepackage{amsfonts}
\usepackage{amssymb,ulem}
\usepackage{color}%
\usepackage{dcolumn}

\usepackage{MnSymbol,wasysym}
\usepackage{braket}
\usepackage{eurosym}
\usepackage{calrsfs}
\usepackage{multirow}
\usepackage{float}
\usepackage[usenames,dvipsnames,svgnames]{xcolor}

\newcommand{\RNum}[1]{\uppercase\expandafter{\romannumeral #1\relax}}
\usepackage[colorlinks=true,linkcolor=blue,urlcolor=blue,filecolor=black,citecolor=red,pdfstartview=FitV,pdftitle={},pdfsubject={},pdfkeywords={},pdfpagemode=None,bookmarksopen=true]{hyperref}
\usepackage[title]{appendix}
\makeatletter
\newcommand*{\rom}[1]{\expandafter\@slowromancap\romannumeral #1@}
\makeatother

\begin{document}
\baselineskip=0.5 cm

\title{Scalar field perturbation around a rotating hairy  black hole:
quasinormal modes, quasibound states and  superradiant  instability}

\author{Yun-He Lei}
\email{leiyunhe2022@163.com}
\affiliation{Center for Gravitation and Cosmology, College of Physical Science and Technology, Yangzhou University, Yangzhou, 225009, China}

\author{Zhen-Hao Yang}
\email{yangzhenhao$_$yzu@163.com}
\affiliation{Center for Gravitation and Cosmology, College of Physical Science and Technology, Yangzhou University, Yangzhou, 225009, China}

\author{Xiao-Mei Kuang}
\email{xmeikuang@yzu.edu.cn (corresponding author)}
\affiliation{Center for Gravitation and Cosmology, College of Physical Science and Technology, Yangzhou University, Yangzhou, 225009, China}

\begin{abstract}
\baselineskip=0.5 cm
We consider the quasinormal modes, quasibound states and superradiant instability of a rotating hairy  black hole, which possesses a Horndeski hair as deviation from Kerr black hole, under the perturbation of massive scalar field. With the use of the matrix method, we mainly calculate the eigenfrequencies related to those modes of the perturbation. Under the perturbation of the massless scalar field,  the Horndeski hair and spin parameter have significant influences on the quasinormal frequency, but its imaginary part is always finite negative and no unstable mode is found. Under the perturbation of the massive scalar field, we focus on the eigenfrequencies of quasibound states and find the modes of which the imaginary part of eigenfrequencies is positive, indicating that the black hole undergoes superradiant instability. Then we scan the parameters and figure out a diagram in the space of Horndeski hair and spin parameters to distinguish the rotating hairy  black hole with superradiant instability from the stable one.
\end{abstract}

\maketitle
\tableofcontents
\newpage
\section{Introduction}\label{sec 1}

The stability of black hole solution under certain perturbations  is an over-lasting crucial problem in black hole physics.  Quasinormal modes (QNMs) with  specific eigenfrequencies (Quasinormal frequencies, or QNFs)  play a remarkable role in this scenario.   QNFs of a field perturbation on the black hole are infinite discrete spectrum of complex frequencies,  of which the real part determines the oscillation timescale of the QNMs, while the imaginary part determines their exponential decaying timescale. It is one of the most important characteristics of black hole geometry, and the interest of QNMs in more fundamental physics can be referred into the reviews \cite{Nollert:1999ji,Berti:2009kk,Konoplya:2011qq}.  Early in the 1950s, Regge and Wheeler studied the stability of black holes in general relativity (GR) by analyzing QNFs of gravitational perturbation in Schwarzschild black hole, and they found that the perturbation field  oscillates and decays over time such that the
Schwarzschild black hole is stable \cite{Regge:1957td}. Later, various external  field perturbations in the Schwarzschild black hole were explored and further verified the stability of the Schwarzschild black hole in GR \cite{Konoplya:2003ii,Cardoso:2003vt,Cho:2003qe,Jing:2005dt} .


However, rotating neutral black holes(i.e. Kerr black holes) in GR may develop instabilities under certain conditions.  In fact, if an ingoing bosonic wave is scattered off a Kerr black hole with the horizon angular velocity $\Omega_H$, the scattered wave will be amplified once the  superradiance  condition $0<\omega<m\Omega_H$ is satisfied, where $\omega$ is the frequency of the wave and $m$ is the azimuthal number \cite{Starobinsky:1973aij}. The extra energy of the scattered wave is obtained from the rotational energy of the black hole, so this is a wave analogue of the Penrose process \cite{Penrose:1971uk,Misner:1972kx}. In this process, if there exists a hypothetic reflective ``mirror" that makes the amplified wave be scattered back and forth between
the ``mirror" and the black hole, the amplifying  proceeds continuously and  will finally cause the  black hole background to suffer from superradiant instability \cite{Brito:2015oca}. Such a system was also dubbed the ``black hole bomb" \cite{Press:1972zz}.  The Kerr black hole was found to be stable under massless scalar, electromagnetic or gravitational perturbations  according to the analysis of those QNMs \cite{Press:1973zz,Teukolsky:1974yv}. However, once the perturbing  scalar field is massive, Kerr black hole was found to undergo superradiant instability   because the mass term of the perturbing scalar field acts as the natural mirror \cite{Damour:1976kh,Zouros:1979iw,Detweiler:1980uk,Li:2012rx,Dolan:2007mj,Cardoso:2005vk,Hod:2009cp,Damour:1976jd}. Moreover, it was addressed in \cite{Brito:2015oca,Dolan:2007mj}  that such superradiatively unstable mode may be induced from the quasi-bound state (QBS) of the massive scalar field, which  decays exponentially far away from the black hole with certain eigenfrequency.

In GR, the eigenfrequencies for QNM and QBS around a Kerr black hole are only characterized by three parameters: the mass  and  angular momentum of the black hole and the mass of external field.
Though recent observations of gravitational wave generated from binary compact objects \cite{LIGOScientific:2016aoc,LIGOScientific:2018mvr} and black hole shadow \cite{EventHorizonTelescope:2019uob,EventHorizonTelescope:2022wkp} provide us with remarkable  chances to test GR in the strong field regime, the uncertainties in the data leave some space for alternative theories of gravity and non-Kerr metric of  black holes.  In addition, despite the great success, there are still some challenging problems for GR, such as  the explanation of the Universe expansion
history, the large scale structure and the understanding of quantum gravity \cite{SupernovaSearchTeam:1998fmf,Weinberg:1988cp,Carroll:2000fy}. Therefore,  a more general theory of gravity is required  from both observational and theoretical aspects. Plenty  of modified gravitational theories have been proposed \cite{Clifton:2011jh,Bakopoulos:2020mmy,Corelli:2020hvr}, among which  the usual way is to modify the action of GR. The scalar tensor theories, which contains a scalar field in addition to the metric field are considered as the simplest nontrivial extensions of GR. The most general scalar tensor theory with second order equations of motion for both the metric and the scalar field in four dimensions is the Horndeski theory, which is free of the Ostrogradski instabilities \cite{Horndeski:1974wa,Kobayashi:2019hrl} and is found to be
equivalent to the generalized Galileon theory \cite{Deffayet:2011gz,Kobayashi:2011nu} derived in a different way.
The observational constraints or bounds on Horndeski gravity have been extensively explored in \cite{Bellini:2015xja,Bhattacharya:2016naa,Kreisch:2017uet,Hou:2017cjy,SpurioMancini:2019rxy,Allahyari:2020jkn}. Moreover, Horndeski gravity attracts lots of attention in both cosmological and astrophysical communities because it has significant consequences  in describing the accelerated expansion and other interesting features (please see \cite{Kobayashi:2019hrl} for a review), as well as holographic applications \cite{Feng:2015oea,Kuang:2016edj,Jiang:2017imk,Baggioli:2017ojd,Feng:2018sqm,Wang:2019jyw,Zhang:2022hxl,Bravo-Gaete:2020lzs,Bravo-Gaete:2022lno} and references therein.

In particular, the Horndeski framework has also been used to test the no hair theorem of black holes, which states that the isolated black hole in GR is merely characterized by three parameters: mass, electric charge and angular momentum.  Due to the extra scalar field in the action, it is natural to figure out if there are black hole solutions with scalar hair, and this makes sense because similar to GR, the Horndeski theory has diffeomorphism invariance and second order field equations. The no hair theorem for the shift symmetric Horndeski theory \cite{Hui:2012qt} has proved that a static and spherically symmetric black hole cannot sustain a nontrivial scalar field with vanishing boundary condition at infinity. But later it turns out that  there is a loophole in the demonstration and the counter hairy black hole solution was constructed in \cite{Sotiriou:2013qea,Sotiriou:2014pfa}. The no hair theorem has also been considered in \cite{Babichev:2017guv} for the shift symmetric Horndeski and beyond Horndeski theories, which also admit the hairy static and spherically symmetric black hole solutions with nontrivial scalar profile. Up to date, many hairy black holes have been constructed and analyzed in Horndeski theory, including the radially dependent \cite{Rinaldi:2012vy,Cisterna:2014nua,Feng:2015oea} and time--dependent scalar hair \cite{Babichev:2013cya,Babichev:2017lmw,BenAchour:2018dap}, among which the black holes with linear time--dependent scalar hair is unstable under perturbation \cite{Khoury:2020aya}, hence ruling out the possibility of this type of hairy solutions. Fortunately,  a spherically symmetric hairy black hole solution in shift symmetric Horndeski theory has recently been constructed  in \cite{Bergliaffa:2021diw},
which was verified to be stable under various massless external field perturbations by us \cite{Yang:2023lcm}.  Some observational investigations have been carried out for this static hairy Horndeski black hole, for example, the strong gravitational lensing \cite{Kumar:2021cyl} and photon rings in the black hole image \cite{Wang:2023vcv}, which shew  that the Horndeski hair has a significant influence on the related observable.
Especially, with the use of the revised Newman--Janis algorithm \cite{Azreg-Ainou:2014pra,Azreg-Ainou:2014aqa}, the authors of \cite{Walia:2021emv}  soon constructed its axially symmetric counterpart, known as rotating black hole with Horndeski hair. The thermodynamic properties and weak gravitational lensing \cite{Walia:2021emv},  black hole shadow  \cite{Afrin:2021wlj} and  superradiant energy extraction \cite{Jha:2022tdl}  have been extensively studied in this rotating hairy black hole.

In this paper, we will study the QNMs and QBSs for a massive perturbing scalar field on the rotating hairy  black hole in Horndeski theory. Our study motivates from two aspects. One is that the computation of eigenfrequencies of QNMs for scalar field perturbation is a first step to analyze the dynamical stability of the rotating hairy black hole. The other is that the interesting QBSs could be the potential source of  the superradiatively unstable modes as we aforementioned. It is noticed that a brief investigation of superradiant instabilities in the rotating hairy metric was performed in \cite{Jha:2022tdl} using approximate analytical methods. Our work expands the
results of \cite{Jha:2022tdl}  by analyzing in detail  the eigenfrequency of these unstable modes as a function of the black hole parameters.
We will explore how the hair parameter influences this instability apart from the rotation. To this end, we will use the matrix method to numerically determine the frequency spectrums of QNMs, QBSs and superradiatively unstable modes.

This paper is organized as follows. In Sec.\ref{sec 2} we present the background geometry of the rotating hairy Horndeski black hole, derive the master equations of the scalar field perturbation and analyze the corresponding boundary conditions. The eigenfrequencies of QNMs, QBSs and superradiant instabilities of the perturbation are explored in Sec.\ref{sec 3}, in which we mainly analyze the effect of the hair parameter and figure out a `phase' diagram in black hole parameter space. Sec.\ref{sec 4} contains our conclusions.

\section{Background geometry and setup of the scalar field perturbation}\label{sec 2}
\begin{figure*}[htbp]
\centering
\includegraphics[height=5.5cm]{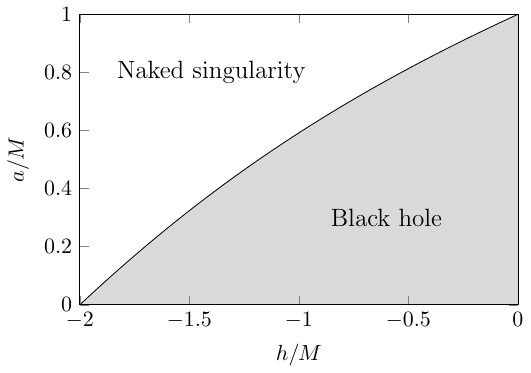}
\caption{\label{BH region}The $(a/M, h/M)$ parameter regions in which the metric \eqref{metric} describes a black hole (gray region) or a naked singularity (white region), the parameters on the black curve that separates these two regions correspond to a extreme black hole.}
\end{figure*}

\begin{figure*}[htbp]
\includegraphics[height=4.5cm]{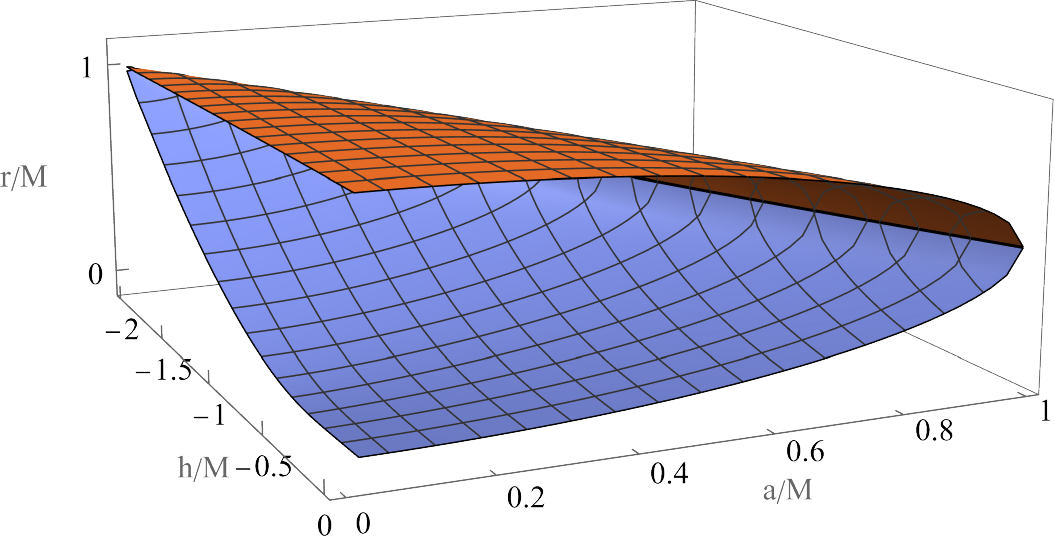}
\includegraphics[height=4.5cm]{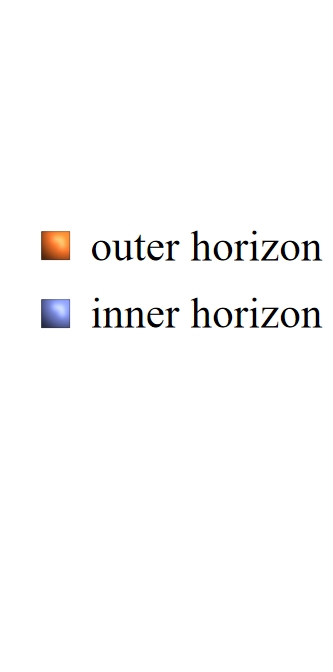}
\caption{\label{BH region2} The horizons $r_\pm$ as functions of the black hole parameters. The orange part indicates the behavior of event horizon $r_{+}$ while the blue part describe the behavior of Cauchy horizon $r_{-}$.
}
\end{figure*}

The action for the  dubbed  quartic Horndeski gravity reads as \cite{Babichev:2017guv}
\begin{eqnarray}\label{eq:action}
S=\int d^4x \sqrt{-g}\big[Q_2+Q_3\Box\phi+Q_4R+Q_{4,\chi}\left((\Box\phi)^2-(\nabla^\mu\nabla^\nu\phi)(\nabla_\mu\nabla_\nu\phi)\right)
+Q_5G_{\mu\nu}\nabla^\mu\nabla^\nu\phi\nonumber\\
-\frac{1}{6}Q_{5,\chi}\left((\Box\phi)^3-3(\Box\phi)(\nabla^\mu\nabla^\nu\phi)(\nabla_\mu\nabla_\nu\phi)
+2(\nabla_\mu\nabla_\nu\phi)(\nabla^\nu\nabla^\gamma\phi)(\nabla_\gamma\nabla^\mu\phi)\right)\big],
\end{eqnarray}
where $\chi=-\partial^\mu\phi\partial_\mu\phi/2$ is the canonical kinetic term, $Q_i~(i=2,3,4,5)$ are arbitrary functions of $\chi$ and $Q_{i,\chi} \equiv \partial Q_{i}/\partial \chi $, $R$ is the Ricci scalar and $G_{\mu\nu}$ is the Einstein tensor. A static hairy black hole in a specific quartic Horndeski theory, saying that $Q_5$ in the above action vanishes, has been constructed in \cite{Bergliaffa:2021diw} with the metric
\begin{eqnarray}\label{eq:metric1}
ds^2=-f(r)dt^2+\frac{dr^2}{f(r)}+r^2(d\theta^2+\sin^2\theta d\varphi^2)~~~
\mathrm{with}~~~f(r)=1-\frac{2M}{r}+\frac{h}{r}\ln\left(\frac{r}{2M}\right).
\end{eqnarray}
Here, $M$ and $h$ are the parameters related to the black hole mass and scalar hair.
The metric reduces to Schwarzschild case as $h\to 0$, and it is asymptotically flat. The hairy black hole described by the above metric was found to be stable under various external field perturbations by computing their QNFs and time evolutions \cite{Yang:2023lcm}. Besides, some observable phenomena affected by the Horndeski hair have also been explored in \cite{Kumar:2021cyl,Wang:2023vcv}. Very recently, the axial symmetric counterpart of metric \eqref{eq:metric1} was constructed, and in the Boyer-Lindquist coordinates the metric of the corresponding rotating hairy black hole reads as \cite{Walia:2021emv}
\begin{equation}\label{metric}
ds^2=-\left[1-\frac{2\tilde{M}(r)r}{\Sigma}\right]dt^2+\frac{\Sigma}{\Delta}dr^2+\Sigma d\theta^2+\frac{(r^2+a^2)^2-a^2\Delta\sin^2\theta}{\Sigma}\sin^2\theta d\varphi^2-\frac{4a\tilde{M}(r)r}{\Sigma}\sin^2\theta dtd\varphi,
\end{equation}
with the functions
\begin{equation}
\Delta=r^2-2\tilde{M}(r)r+a^2,\quad\Sigma=r^2+a^2\cos^2\theta,\quad \tilde{M}(r)=M-\frac{h}{2}\ln\left(\frac{r}{2M}\right),
\end{equation}
where $a$ is the spin parameter. The axially symmetric spacetime is also asymptotically flat and reduces to the Kerr spacetime  as $h\to 0$.
The horizons of the rotating hairy black hole are determined by the real positive  roots of $\Delta=0$, which only exist in a certain parameter region of $(a/M, h/M)$ shown in FIG.\ref{BH region}. In detail, in the gray  parameter region, $\Delta=0$ has two real positive roots $r_\pm$, in which the larger root $r_+$ indicates the event horizon while the smaller one $r_-$ denotes the Cauchy horizon. In the white parameter region, $\Delta=0$ has no real root meaning there is no horizon such that the metric \eqref{metric} describes a naked singularity. And the parameters on the black curve that separates these two regions make the rotating hairy  black hole become extremal with $r_-=r_+$.  The horizons $r_\pm$ as functions of the black hole parameters are explicitly shown in FIG.\ref{BH region2}. More interesting properties, including the Komar conserved quantities, thermodynamics, light deflection, shadow, energy emission and energy extraction, of this rotating hairy black hole can be found in \cite{Walia:2021emv,Afrin:2021wlj,Jha:2022tdl}.

Here in order to analyze the dynamical (in)stability of the rotating hairy black hole  \eqref{metric},  we, as the first step, will introduce the massive scalar field as the probe and study the eigenfrequencies of its master equation.
The dynamic  of a perturbing  scalar field around the  rotating hairy black hole is governed by the Klein--Gordon equation
\begin{equation}\label{KG eq}
(\Box-\mu^2)\Phi=\frac{1}{\sqrt{-g}}\partial_\mu(\sqrt{-g}g^{\mu\nu}\partial_\nu\Phi)-\mu^2\Phi=0,
\end{equation}
where $\mu$ is the mass of the scalar field. Then with the decomposition
\begin{equation}
\Phi(t,r,\theta,\varphi)=R(r)S(\theta)e^{-i\omega t+im\varphi},
\end{equation}
in which $\omega$ is the frequency and $m$ is the azimuthal number, we can separate Eq.\eqref{KG eq}, yielding the radial master equation
\begin{equation}\label{R eq}
\Delta\frac{d}{dr}\left[\Delta\frac{d}{dr}R(r)\right]+[\omega^2(a^2+r^2)^2-\Delta(a^2\omega^2+\mu^2r^2+\lambda)-4a\omega mr\tilde{M}(r)+a^2m^2]R(r)=0,
\end{equation}
and the angular master equation
\begin{equation}\label{S eq}
\frac{d}{du}\left[(1-u^2)\frac{d}{du}S(u)\right]+\left[a^2u^2(\omega^2-\mu^2)-\frac{m^2}{1-u^2}+\lambda\right]S(u)=0,
\end{equation}
where $u=\cos\theta$ and $\lambda$ is the separation constant.

Then, we shall analyze  the physically relevant boundary conditions of Eq.\eqref{R eq} and Eq.\eqref{S eq}. The boundary conditions of $S(u)$ can be directly  obtained from its asymptotic solutions  near the boundaries $u\to\pm1$, which are
\begin{equation}\label{BC S}
S(u)\sim\begin{cases}
(1-u)^{\pm\frac{m}{2}},\quad u\to1\\
(1+u)^{\pm\frac{m}{2}},\quad u\to-1
\end{cases}.
\end{equation}
For the radial equation, by introducing the tortoise coordinate
\begin{equation}
r_*=\int\frac{r^2+a^2}{\Delta}dr,
\end{equation}
and the function transformation
\begin{equation}\label{eq:RtoK}
K(r)=\sqrt{r^2+a^2}R(r),
\end{equation}
we can obtain the equation
\begin{equation}
\frac{d^2}{dr_*^2}K(r)+V(r)K(r)=0,
\end{equation}
with
\begin{equation}
V(r)=\left(\omega-\frac{ma}{r^2+a^2}\right)^2-\frac{\Delta}{(r^2+a^2)^2}\left[a^2\omega^2-2am\omega+\mu^2r^2+\lambda+\frac{(a^2-2r^2)\Delta+r(a^2+r^2)d\Delta/dr}{(r^2+a^2)^2}\right].
\end{equation}
It is straightforward to get the asymptotic behavior of $V(r)$ as
\begin{equation}
V(r\to r_+)=(\omega-m\Omega_H)^2 ~~~ \text{and} ~~~V(r\to\infty)=\omega^2-\mu^2,
\end{equation}
with $\Omega_H=\frac{a}{r_+^2+a^2}$ the horizon angular velocity, based on which we can easily obtain the asymptotic solutions as the boundary conditions
\begin{equation}\label{eq:BC_rs}
K(r\to r_+)\sim e^{\pm i\bar{\omega}r_*}~~~ \text{and}~~~K(r\to\infty)\sim e^{\pm ikr_*}.
\end{equation}
with $\bar{\omega}=\omega-m\Omega_H$ and $k=\sqrt{\omega^2-\mu^2}$. Reminding  that  the perturbation cannot come out from the event horizon, we only consider the case with  the minus sign in the first formula of Eq.\eqref{eq:BC_rs}, i.e., the ingoing boundary condition near the event horizon. For the second formula of Eq.\eqref{eq:BC_rs}, we take the plus (minus) sign and ensure $Re(k)>0$ ($Im(k)<0$) to get the purely outgoing (exponentially decaying) boundary condition at the spatial infinity for the QNMs (QBSs) \cite{Dolan:2007mj}. The eigenfrequencies $\omega$ obtained from the master equations are discrete complex numbers, of which the real part describes the oscillation while the imaginary part can be employed to judge the (in)stability of the black hole under the perturbations. QNMs are known to have negative $Im(\omega)$, which means that the perturbation decays and the lifetime is evaluated by $\tau=|Im(\omega)|^{-1}$. But the QBSs may become the source of  the unstable modes with positive $Im(\omega)$, implying an exponentially growing amplitude due to the superradiant instability. Therefore, we shall solve the master equations to obtain the eigenfrequencies, and analyze the possible modes of the perturbation in the rotating hairy black hole.


Since it's not possible to solve the  master equations analytically, so we shall solve them numerically with the use of the matrix method  \cite{Lin:2016sch,Lin:2017oag,Lin:2019mmf} and extract  the eigenfrequencies of the QNMs and QBSs. For the convenience to employ this method, we shall make some transformations to Eq.\eqref{R eq} and Eq.\eqref{S eq} beforehand. Firstly, by rewriting the boundary conditions of the radial equation Eq.\eqref{eq:BC_rs} into the form of function $R$ via Eq.\eqref{eq:RtoK}, and expanding them at the corresponding asymptotic region, we can obtain
\begin{equation}\label{eq:BC_r}
R(r\to r_+)\sim(r-r_+)^{-i\sigma}~~~\text{and}~~~ R(r\to\infty)\sim e^{\pm ikr\mp \frac{1}{2}ihk(\ln\frac{r}{2M})^2}r^{-1\pm2iMk},
\end{equation}
where $\sigma=\frac{r_+(r_+^2+a^2)}{r_+(r_++h)-a^2}\bar{\omega}$, and the upper (lower) signs  correspond to the plus (minus) sign in the second formula of Eq.\eqref{eq:BC_rs}. Then, in order to factor out the asymptotic behaviors of the radial and angular functions, we redefine
\begin{equation}
\begin{split}
R(r)&=(r-r_+)^{-i\sigma}r^{i\sigma}e^{\pm ikr\mp \frac{1}{2}ihk(\ln\frac{r}{2M})^2}r^{-1\pm2iMk}\tilde{R}(r),\\
S(u)&=(1-u)^{\frac{m}{2}}(1+u)^{\frac{m}{2}}\tilde{S}(u),
\end{split}
\end{equation}
so that $R(r)$ and $S(u)$ will automatically satisfy the aforementioned boundary conditions. Next, we introduce the new coordinates
\begin{equation}
x=\begin{cases}
1-(\frac{r_+}{r})^{1/3}\quad\text{for QNMs}\\
1-\frac{r_+}{r}\quad\text{for QBSs}
\end{cases}~~~\text{and}~~~
v=\frac{1+u}{2},
\end{equation}
to bring the integration domains into $x\in[0,1]$ and $v\in[0,1]$ to replace  $r\in[r_+,\infty)$ and $u\in[-1,1]$, respectively. Finally, by further performing the transformation
\begin{equation}
\chi_R(x)=\begin{cases}
x(1-x)\tilde{R}(x)\quad\text{for QNMs}\\
x(1-x)^6\tilde{R}(x)\quad\text{for QBSs}
\end{cases}~~~\text{and}~~~
\chi_S(v)=v(1-v)\tilde{S}(v),
\end{equation}
we obtain the equations of $\chi_R$ and $\chi_S$, which we can directly solve by using the matrix method, as
\begin{align}
\mathcal{C}_{R2}(x)\chi_R''(x)+\mathcal{C}_{R1}(x)\chi_R'(x)+\mathcal{C}_{R0}(x)\chi_R(x)=0,\label{chiR}\\
\mathcal{C}_{S2}(v)\chi_S''(v)+\mathcal{C}_{S1}(v)\chi_S'(v)+\mathcal{C}_{S0}(v)\chi_S(v)=0,\label{chiS}
\end{align}
where the coefficient functions $\mathcal{C}_{Ri}$ and $\mathcal{C}_{Si}$ ($i=0\sim2$) are functions of $\omega$ and $\lambda$. Moreover, the corresponding boundary conditions are then converted into the homogeneous form
\begin{align}
\chi_R(0)=\chi_R(1)=0,\label{BC chiR}\\
\chi_S(0)=\chi_S(1)=0.\label{BC chiS}
\end{align}
The last step is to follow the matrix method algorithm  to extract the eigenfrequencies of QNMs and QBSs  by numerically solving the  algebraic equations
\begin{equation}\label{det}
det[\mathcal{M}_R(\omega,\lambda)]=0,\quad\text{and}\quad det[\mathcal{M}_S(\omega,\lambda)]=0,
\end{equation}
where the expressions of the matrix $\mathcal{M}_{R~\text{or}~S}(\omega,\lambda)$ are determined by the coefficient functions in the equations \eqref{chiR} and \eqref{chiS}. Here we prefer not to  repeat the steps of the matrix method algorithm, and readers can refer to  \cite{Lin:2016sch,Lin:2017oag,Lin:2019mmf} for more details.

\section{Results of the perturbing modes and their eigenfrequencies}\label{sec 3}
In this section, we shall analyze the results of QNMs, QBSs and superradiant instabilities of the perturbing scalar field on the rotating hairy black hole background \eqref{metric}. It is noted that all the physical quantities in this section are rescaled  by the mass parameter  $M$ into be dimensionless.

\subsection{Quasinormal modes}
\begin{figure*}[htbp]
\centering
\includegraphics[height=6.5cm]{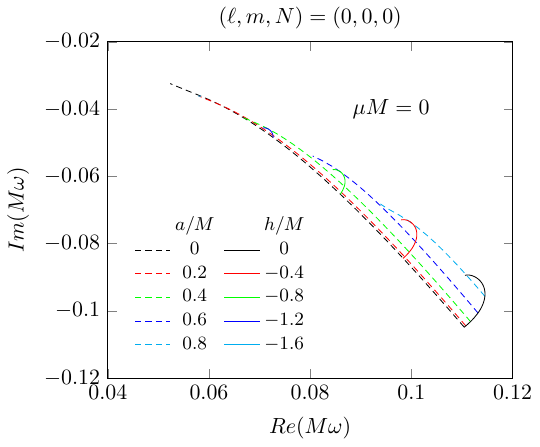}
\caption{\label{QNM l=0}The variation of the real and imaginary parts of eigenfrequencies for fundamental $\ell=m=0$ QNM of the massless scalar field on the rotating hairy Horndeski black hole in terms of the hair (spin) parameter $h/M$ ($a/M$) with several selected spin (hair) parameters. For each curve with given $a/M$ ($h/M$), the value of $h/M$ ($a/M$) starts from $0$ (Kerr case) [(static case)] and then extends toward the negative (positive) direction until the black hole becomes near extreme.}
\end{figure*}

We first consider the dynamic behavior of a massless ($\mu M=0$) scalar field perturbing around the rotating hairy Horndeski black hole. We mainly focus on the effects of spin parameter $a/M$ and the Horndeski hair parameter $h/M$ on the fundamental QNMs with overtone number $N=0$, so we select samples of the two parameters to show the dependence of the fundamental QNFs on them. The fundamental QNFs with the mode $\ell=m=0$ are shown in FIG.\ref{QNM l=0}.  With fixed $a/M$ (see the dashed curves),   as the hair parameter $h/M$ decreases, the real part of the QNFs decreases while the imaginary part increases. This means that comparing to the Kerr black hole, the Horndeski hair makes the massless scalar field perturbation oscillate slower and decay slower.  With fixed $h/M$ (see the solid curves), the effect of the spin parameter on QNFs could depend on the value of $h/M$. In detail, for weaker Horndeski hair, i.e., small $\mid h/M\mid$, as $a/M$ increases, the real part of QNFs first increases and  then decreases which is similar to that found in Kerr black hole \cite{Dolan:2007mj},  but for large enough $\mid h/M\mid$ it decreases monotonically. Meanwhile, the imaginary part first increases with the increasing of $a/M$ and then decreases slightly as the black hole approaches the extreme for all the selected values of $h/M$,  but it tends to a finite negative value for the extremal case. This implies that for the rotating hairy black hole with small $\mid h/M\mid$, the faster spin  makes the corresponding perturbation  oscillate first faster and then slower as the black hole becomes near extremal, similar to that occurs in Kerr black hole. However, in the rotating hairy black hole with large enough $\mid h/M\mid$, the oscillation always becomes slower as $a/M$ increases, this is because the strong suppression effect of $h/M$ on the real part could balance the enhancement effect of $a/M$ and even dominates when the black hole closes to extreme.

\begin{figure*}[htbp]
\centering
\includegraphics[height=4.5cm]{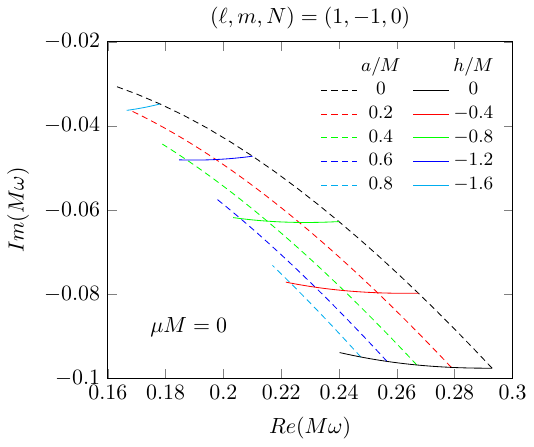}
\includegraphics[height=4.5cm]{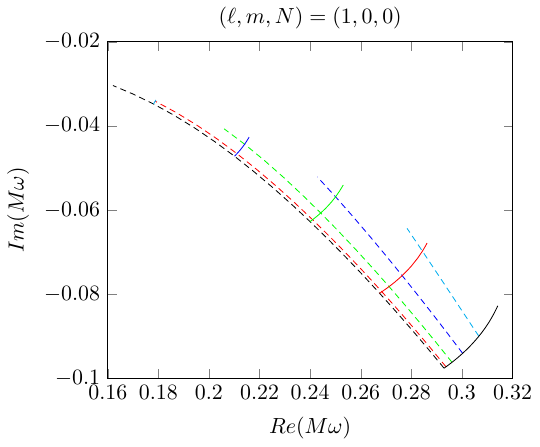}
\includegraphics[height=4.5cm]{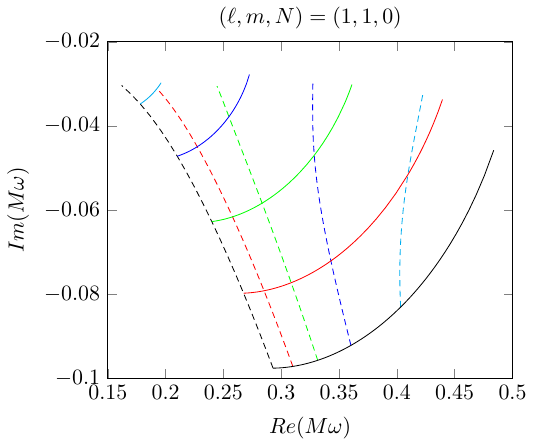}
\caption{\label{QNM l=1} The fundamental QNFs for the modes $\ell=1$, with $m=-1$ (left panel), $m=0$ (middle panel) and $m=1$ (right panel). The setting of black hole parameters are the same as those in  FIG.\ref{QNM l=0}. }
\end{figure*}

To compare the influences of different angular quantum number $\ell$ and azimuthal number $m$ on QNMs, we compute the fundamental QNFs  for $\ell=1$. The results are depicted in FIG. \ref{QNM l=1}, where we exhibit how the QNFs of the modes  $m=-1$ (left panel), $m=0$ (middle panel) and $m=1$ (right panel) are affected by $h/M$ and $a/M$ respectively, with the same selected values of parameters as those in $\ell=0$ case. We can extract the following features. (i) As $h/M$ decreases, the real (imaginary) part of the fundamental QNFs decreases (increases) with any selected $a/M$ for all the $\ell=1$ modes, except for the mode $\ell=m=1$ with large $a/M$ of which the real part first decreases then increases with the decreasing of $h/M$. (ii) The effect of varying $a/M$ can also be read off. For the modes $m=0$ and $m=1$, both the real and the imaginary parts of the QNFs increase as the spin becomes faster for the selected $h/M$. For the mode $m=-1$, as the rotation increases, the real part decreases monotonically for any selected $h/M$, but the imaginary part increases for small $\mid h/M\mid$ and decreases for large $\mid h/M\mid$ on the whole. (iii) According to our study, all the imaginary part of eigenfrequencies are finite negative value and we do not find any unstable mode.

\subsection{Quasibound states and superradiant instabilities}
\begin{figure*}[h]
\centering
\includegraphics[height=6.5cm]{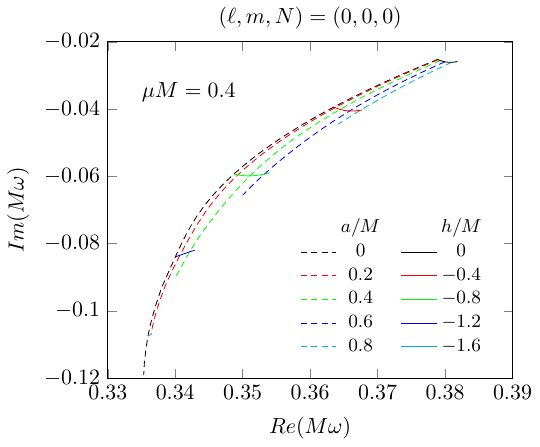}
\caption{\label{QBS l=0} The fundamental eigenfrequencies of  the quasibound states in the mode $\ell=m=0$ of perturbing scalar field with the mass $\mu M=0.4$. The setting of black hole parameters are the same as those in QNMs.}
\end{figure*}

\begin{figure*}[h]
\centering
\includegraphics[height=4.5cm]{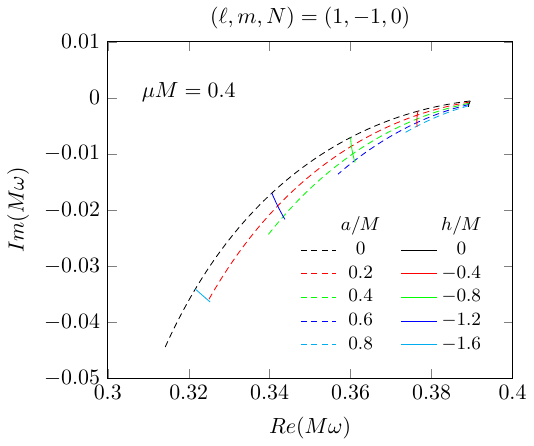}
\includegraphics[height=4.5cm]{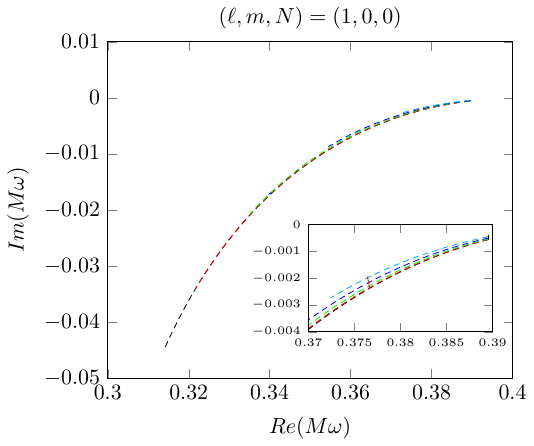}
\includegraphics[height=4.5cm]{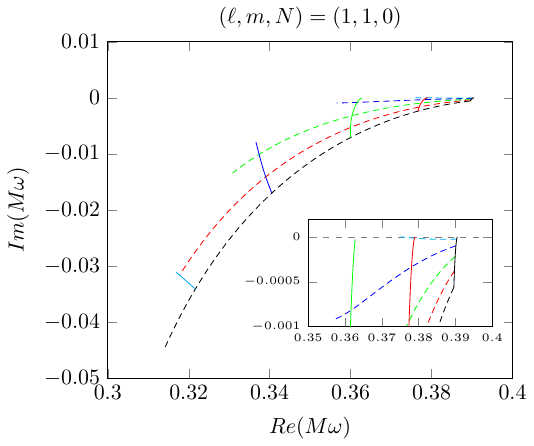}
\caption{\label{QBS l=1}  The fundamental eigenfrequencies of  the quasibound states in the modes $\ell=1$ of perturbing scalar field with the mass $\mu M=0.4$. From left to right, we have $m=-1$, $m=0$ and $m=1$, respectively. Again, the setting of black hole parameters are the same as those in previous studies.}
\end{figure*}

We then turn on the mass of the scalar field and study the existence of QBSs and their eigenfrequencies. As we mentioned in Sec.\ref{sec 1}, the mass term may trigger the superradiant instability that we will show in this subsection. To be specific, we fix the mass of the scalar field  $\mu M=0.4$ as it was done in \cite{Siqueira:2022tbc}, and  compute the fundamental eigenfrequencies of QBSs for the modes $\ell=0$ and $\ell=1$ for the same selected parameters as those in the QNMs. The results are depicted in FIG.\ref{QBS l=0} and FIG.\ref{QBS l=1}, from which we can find some common features of the modes. (i) The real part of the frequencies exhibits a typical feature of the QBSs, i.e., they are always smaller than the the mass of the scalar field. (ii) Similar to the QNFs, the strength of the Horndeski hair also influences the effect of the spin parameter on the QBS frequency, which could be different for different modes. Here we will pay more attention to the effect of Horndeski hair on the eigenfrequencies rather than starchily setting out the effect of rotation.
\begin{figure*}[h]
\centering
\includegraphics[height=6.4cm]{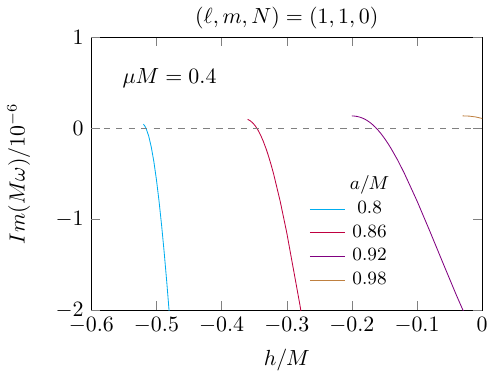}
\includegraphics[height=6.4cm]{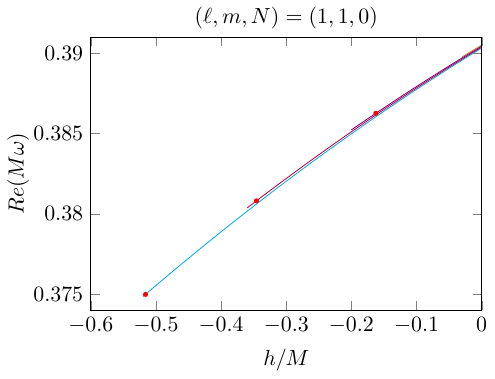}
\caption{\label{fig:srist_varying_a} The fundamental eigenfrequency of the QBSs in the mode  $\ell=m=1$ of massive ($\mu M=0.4$) scalar field as a function of $h/M$ with several selected $a/M$. In the plots, each curve for fixed $a/M$ extends toward the negative $h/M$ direction until the black hole becomes near extreme. In the left plot, there exists positive values of the imaginary part, indicating the superradiant instabilities. In the right plot, the solid curves denote the numerical results of the real part. The red dots denote the eigenfrequencies of the bound state or scalar clouds for which the imaginary part vanishes, so the parameters are extracted from the intersections between each curve and the dashed line in the left plot.
The real part of these bound states marked by red dots are directly evaluated by $\omega_{sc}=m\Omega_H$.}
\end{figure*}

In FIG.\ref{QBS l=0} with $\ell=0$, the imaginary part of the eigenfrequency is always negative, indicating that the mode is stable. With a fixed spin parameter, the stronger Horndeski hair suppresses both the real and imaginary parts of the eigenfrequency, which is different from the one that happens in QNFs.  Moreover, it is obvious that by comparing FIG.\ref{QBS l=0} and FIG.\ref{QNM l=0}, the real part of the QBSs of the massive scalar field with $\mu M=0.4$ is larger than those of the QNMs of the massless scalar field, indicating faster oscillation for the former. In FIG.\ref{QBS l=1} with $\ell=1$,  with the selected $a/M$, the real and the imaginary parts of QBSs from all modes always decrease with the decreasing of $h/M$, however, the inset of the right panel shows the exception for the mode $\ell=m=1$ with $a/M=0.8$, where the imaginary part first decreases slightly then increases and eventually becomes positive. These QBSs  with positive $Im(\omega)$ means that the amplitude of these states will grow exponentially over time, leading to the superradiant instabilities. Moreover, note that the superradiant instabilities only exist in the $m>0$ states, as the result of the superradiance condition $0<\omega<m\Omega_H$.

\begin{figure*}[h]
\centering
\includegraphics[height=6.1cm]{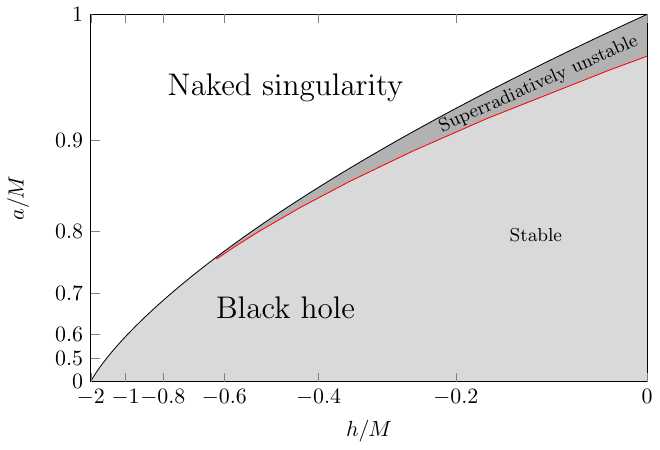}
\caption{\label{fig:phase diagram} The parameter regions for the stable rotating hairy Horndeski black hole and the one with superradiant instability under the perturbation of massive ($\mu M=0.4$) scalar field, of the fundamental $\ell=m=1$ QBSs.  The black curve corresponds to the extreme black hole as in FIG. \ref{BH region}, and the red curve corresponds to the onset of superradiant instability, or bound state or scalar cloud with the eigenfrequency $\omega_{sc}=m\Omega_H$ which is purely real.}
\end{figure*}
\begin{figure*}[h]
\centering
\includegraphics[height=6.4cm]{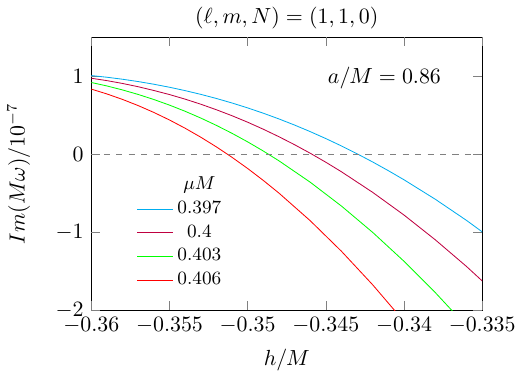}
\includegraphics[height=6.4cm]{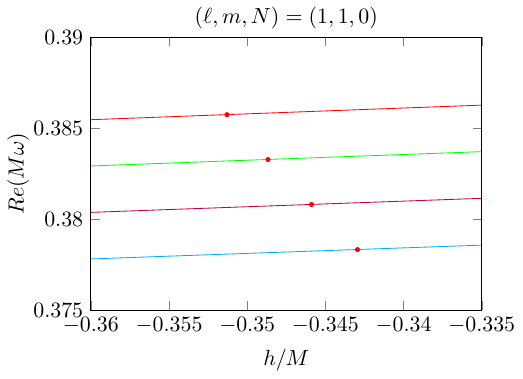}
\caption{\label{fig:srist_varying_mu}The fundamental eigenfrequency of the   QBSs in the mode $\ell=m=1$ as a function of $h/M$ for selected masses of the scalar field. Here we fix $a/M=0.86$. In the left plot for imaginary part, we see that the onset of superradiant instability is very sensitive to the scalar mass. Again, the red dots in the right plot denote the eigenfrequencies of bound state or scalar cloud, which we evaluate via $\omega_{sc}=m\Omega_H$.}
\end{figure*}

To clearly show how the parameters of hairy black hole  affect the superradiant instabilities, we further plot the eigenfrequencies  as a function of the Horndeski hair for several large spin parameters in FIG. \ref{fig:srist_varying_a}. In the left plot, the dashed line with $Im(M \omega)=0$ separates the decaying modes with $Im(M \omega)<0$ from the superradiatively unstable modes with $Im(M \omega)>0$. In fact, the modes with $Im(M \omega)=0$ are bound states which could form as the scalar cloud around the rotating hairy black hole. As addressed in \cite{Hod:2012px}, the eigenfrequency of the scalar cloud is purely real $\omega_{sc}=m\Omega_H$, which means that the scalar cloud is synchronized with the rotation of the hairy black hole such that there is no energy flux towards the black hole \cite{Hod:2012px,Herdeiro:2014goa,Benone:2014ssa}. In the right plot, our numerical real part matches well with $\omega_{sc}=m\Omega_H$ marked by the red dots for the scalar cloud. In a word, the Horndeski hair significantly affects the superradiant instability and will make the onset of superradiant instability need a smaller spin parameter. We collect the data for the scalar cloud and superradiatively unstable modes into the $(a/M,h/M)$ space, depicted in FIG.\ref{fig:phase diagram} where we rescale the axes, and the red curve denotes the onset of instability or the black hole with a scalar cloud. The rotating hairy Horndeski black hole with the parameters in the light gray region is stable under the perturbation of the scalar field with mass $\mu M=0.4$, while the black hole with the parameters  in the dark gray region could undergo superradiant instability.  It is shown that comparing to the Kerr black hole $(h=0)$, the superradiant instability can occur for slower rotating hairy Horndeski black hole but the parameter region of the existence of superradiant instability is narrower.

Before closing  this section, we shall study the effect of the scalar field mass $\mu M$ on the QBSs and the superradiant instabilities.  As addressed in \cite{Dolan:2007mj}, the superradiant instability of Kerr black hole occurs when the mass is in the range of  $0< \mu M < \mu_c M$, where $\mu_c M$ depends on the spin of black hole and we name it as the upper bound of the scalar mass corresponding to $Im(M \omega)=0$.   Here we fix the spin parameter $a/M=0.86$ and calculate the eigenfrequencies of the fundamental $\ell=m=1$ mode as a function of the Horndeski hair. Due to the technical problems, it is difficult for us to scan all the values of $\mu M$, so we only give the results with  selected values of $\mu M$ in FIG. \ref{fig:srist_varying_mu} .
The figure shows that in the selected parameter region, as $\mu M$ increases, the imaginary (real) part of the eigenfrequencies decreases (increases). Moreover,   for the rotating  hairy Horndeski black hole with a smaller $h/M$, the upper bound  of  $\mu M$ for  superradiant instability increases. This suggests that comparing to the Kerr black hole, the mass of the scalar cloud for the rotating  hairy Horndeski black hole is enhanced. In addition, we evaluate the eigenfrequencies of the scalar cloud via  $\omega_{sc}=m\Omega_H$ marked by the red dots in the right plot, which are coincident with the numerical results. It is shown that the scalar cloud with larger mass oscillates faster in the rotating hairy Horndeski black hole.

\section{Conclusions}\label{sec 4}

Neutral black holes in general relativity are known to be characterized by two hairs: the mass and spinning parameter. However, theories of gravity beyond GR usually allow additional hairs, which in  general modify various properties of the spacetime. In this paper, with the use of matrix method, we investigated the eigenfrequencies associated with the perturbation of a massive scalar field around a rotating hairy black hole in Horndeski theory.

Under the perturbation of massless scalar field, we found that comparing to the Kerr black hole with the same fixed spin parameter, the Horndeski hair could make the perturbation mode $\ell=m=0$ oscillate slower and decay slower, corresponding to smaller real part and larger imaginary part of the QNFs. The effect of the spin parameter on the QNFs could depend on the fixed value of $h/M$. For small $\mid h/M\mid$, the real part of QNFs first increases and then decreases with the increase of $a/M$, which is similar to that found in Kerr black hole. While for large enough $\mid h/M\mid$, it decreases monotonically, which is understandable because the strong suppression effect of $\mid h/M\mid$ on the real part could balance the enhancement effect of $a/M$ and even dominates when the black hole closes to the extreme. Meanwhile, for all the selected values of $h/M$, the imaginary part first increases with the increasing of $a/M$ but then decreases slightly as the black hole closes to the extreme. In addition, QNFs with $\ell=1$ and $m=-1,0, 1$ were also calculated. The Horndeski hair also has a significant influence on those QNFs, though the rules may be different from those for the mode $\ell=m=0$. Nevertheless, all the QNFs for various modes have finite negative imaginary part, which in some sense suggests that the rotating hairy Horndeski black hole is stable under the perturbation of a massless scalar field.

Then we turned on the mass of the scalar field perturbation, and mainly checked the eigenfrequencies of the quasibound state. For the scalar mass $\mu M=0.4$, we found significant effects of the Horndeski hair and spin parameter on the eigenfrequencies of the quasibound states with $\ell=0$ and $\ell=1$. Especially, the stronger Horndeski hair always suppresses both the real and imaginary parts of the eigenfrequency for all the modes, except for the mode $\ell=m=1$ with $a/M=0.8$ of which the imaginary part first decreases slightly and then increases to be positive. This could be associated with the superradiant instability of the black hole under the perturbed mode $\ell=m=1$. Afterward, we scanned the parameter space to fix the onset of superradiant instability, also known as bound state or scalar cloud, which satisfies $Im(M\omega)=0$ and $Re(M \omega)=m \Omega_H$. Our results indicate that due to the existence of the Horndeski hair, the rotating hairy black hole with a smaller spin than Kerr black hole could undergo superradiant instability. Finally, we discussed the effect of mass of the scalar field on the quasibound state and superradiant instability. Our study in the selected parameter region, shew that the upper bound of the scalar field  mass for the superradiant instability increases when the value $h/M$ is smaller. This implies that comparing to the Kerr black hole, the range of the scalar field mass for superradiant instability of the rotating hairy Horndeski black hole may be extended.

To conclude, this study provided the preliminary step to analyze the (in)stability of the rotating hairy black hole in Horndeski theory, which we believe deserves to extend into the perturbations of fields with higher spin. Our findings on superradiant instability expand the approximate analytical results of \cite{Jha:2022tdl}. Moreover, our results rich the phenomenal properties introduced by the Horndeski hair, which we hope to give some insight in the test of the no-hair theorem of black holes as well as the observational applications of hairy black holes.

\begin{acknowledgments}
This work is partly supported by Natural Science Foundation of China under Grants No. 12375054, Natural Science Foundation of Jiangsu Province under Grant No.BK20211601 and the Postgraduate Research \& Practice Innovation Program of Jiangsu Province under Grant No. KYCX23\_3501.
\end{acknowledgments}

\bibliography{ref}
\bibliographystyle{apsrev}

\end{document}